
\input harvmac 
%
\newif\ifdraft

\noblackbox
\catcode`\@=11
\newif\iffrontpage
%
\ifx\answ\bigans
\def\titleft{\titsm}
\magnification=1200\baselineskip=15pt plus 2pt minus 1pt
%
\advance\hoffset by-0.075truein
\hsize=6.15truein\vsize=600.truept\hsbody=\hsize\hstitle=\hsize
\else\let\lr=L
\def\titleft{\titla}
\magnification=1000\baselineskip=14pt plus 2pt minus 1pt
%
\vsize=6.5truein
\hstitle=8truein\hsbody=4.75truein
\fullhsize=10truein\hsize=\hsbody
\fi
\parskip=4pt plus 15pt minus 1pt

\font\titla=cmr10 scaled\magstep3
\font\tenmss=cmss10
\font\absmss=cmss10 scaled\magstep1
\newfam\mssfam
\font\footrm=cmr8  \font\footrms=cmr5
\font\footrmss=cmr5   \font\footi=cmmi8
\font\footis=cmmi5   \font\footiss=cmmi5
\font\footsy=cmsy8   \font\footsys=cmsy5
\font\footsyss=cmsy5   \font\footbf=cmbx8
\font\footmss=cmss8
\def\footfont{\def\rm{\fam0\footrm}
\textfont0=\footrm \scriptfont0=\footrms
\scriptscriptfont0=\footrmss
\textfont1=\footi \scriptfont1=\footis
\scriptscriptfont1=\footiss
\textfont2=\footsy \scriptfont2=\footsys
\scriptscriptfont2=\footsyss
\textfont\itfam=\footi \def\it{\fam\itfam\footi}
\textfont\mssfam=\footmss \def\mss{\fam\mssfam\footmss}
\textfont\bffam=\footbf \def\bf{\fam\bffam\footbf} \rm}
\def\tenpoint{\def\rm{\fam0\tenrm}
\textfont0=\tenrm \scriptfont0=\sevenrm
\scriptscriptfont0=\fiverm
\textfont1=\teni  \scriptfont1=\seveni
\scriptscriptfont1=\fivei
\textfont2=\tensy \scriptfont2=\sevensy
\scriptscriptfont2=\fivesy
\textfont\itfam=\tenit \def\it{\fam\itfam\tenit}
\textfont\mssfam=\tenmss \def\mss{\fam\mssfam\tenmss}
\textfont\bffam=\tenbf \def\bf{\fam\bffam\tenbf} \rm}
\ifx\answ\bigans\def\abstractfont{\tenpoint}\else
\def\abstractfont{\def\rm{\fam0\absrm}
\textfont0=\absrm \scriptfont0=\absrms
\scriptscriptfont0=\absrmss
\textfont1=\absi \scriptfont1=\absis
\scriptscriptfont1=\absiss
\textfont2=\abssy \scriptfont2=\abssys
\scriptscriptfont2=\abssyss
\textfont\itfam=\bigit \def\it{\fam\itfam\bigit}
\textfont\mssfam=\absmss \def\mss{\fam\mssfam\absmss}
\textfont\bffam=\absbf \def\bf{\fam\bffam\absbf}\rm}\fi
%
\def\f@@t{\baselineskip10pt\lineskip0pt\lineskiplimit0pt
\bgroup\aftergroup\@foot\let\next}
\setbox\strutbox=\hbox{\vrule height 8.pt depth 3.5pt width\z@}
\def\vfootnote#1{\insert\footins\bgroup
\baselineskip10pt\footfont
\interlinepenalty=\interfootnotelinepenalty
\floatingpenalty=20000
\splittopskip=\ht\strutbox \boxmaxdepth=\dp\strutbox
\leftskip=24pt \rightskip=\z@skip
\parindent=12pt \parfillskip=0pt plus 1fil
\spaceskip=\z@skip \xspaceskip=\z@skip
\Textindent{$#1$}\footstrut\futurelet\next\fo@t}
\def\Textindent#1{\noindent\llap{#1\enspace}\ignorespaces}
\def\footnote#1{\attach{#1}\vfootnote{#1}}%

\def\foot{\attach\footsymbolgen\vfootnote{\footsymbol}}
\let\footsymbol=\star
\newcount\lastf@@t           \lastf@@t=-1
\newcount\footsymbolcount    \footsymbolcount=0
\def\footsymbolgen{\relax\footsym
\global\lastf@@t=\pageno\footsymbol}
\def\footsym{\ifnum\footsymbolcount<0
\global\footsymbolcount=0\fi
{\iffrontpage \else \advance\lastf@@t by 1 \fi
\ifnum\lastf@@t<\pageno \global\footsymbolcount=0
\else \global\advance\footsymbolcount by 1 \fi }
\ifcase\footsymbolcount \fd@f\star\or
\fd@f\dagger\or \fd@f\ast\or
\fd@f\ddagger\or \fd@f\natural\or
\fd@f\diamond\or \fd@f\bullet\or
\fd@f\nabla\else \fd@f\dagger
\global\footsymbolcount=0 \fi }
\def\fd@f#1{\xdef\footsymbol{#1}}
\def\space@ver#1{\let\@sf=\empty \ifmmode #1\else \ifhmode
\edef\@sf{\spacefactor=\the\spacefactor}
\unskip${}#1$\relax\fi\fi}
\def\attach#1{\space@ver{\strut^{\mkern 2mu #1}}\@sf}
%
\newif\ifnref
\def\rrr#1#2{\relax\ifnref\nref#1{#2}\else\ref#1{#2}\fi}
\def\ldf#1#2{\begingroup\obeylines
\gdef#1{\rrr{#1}{#2}}\endgroup\unskip}
\def\nrf#1{\nreftrue{#1}\nreffalse}
\def\doubref#1#2{\refs{{#1},{#2}}}
\def\multref#1#2#3{\nrf{#1#2#3}\refs{#1{--}#3}}
\nreffalse
\def\refout{\listrefs}
%
\def\eqn#1{\xdef #1{(\secsym\the\meqno)}
\writedef{#1\leftbracket#1}%
\global\advance\meqno by1\eqno#1\eqlabeL#1}
\def\eqnalign#1{\xdef #1{(\secsym\the\meqno)}
\writedef{#1\leftbracket#1}%
\global\advance\meqno by1#1\eqlabeL{#1}}
%
\def\chap#1{\newsec{#1}}
\def\chapter#1{\chap{#1}}
\def\sect#1{\subsec{{ #1}}}
\def\section#1{\sect{#1}}
\def\\{\ifnum\lastpenalty=-10000\relax
\else\hfil\penalty-10000\fi\ignorespaces}
\def\note#1{\leavevmode%
\edef\@@marginsf{\spacefactor=\the\spacefactor\relax}%
\ifdraft\strut\vadjust{%
\hbox to0pt{\hskip\hsize%
\ifx\answ\bigans\hskip.1in\else\hskip .1in\fi%
\vbox to0pt{\vskip-\dp
\strutbox\sevenbf\baselineskip=8pt plus 1pt minus 1pt%
\ifx\answ\bigans\hsize=.7in\else\hsize=.35in\fi%
\tolerance=5000 \hbadness=5000%
\leftskip=0pt \rightskip=0pt \everypar={}%
\raggedright\parskip=0pt \parindent=0pt%
\vskip-\ht\strutbox\noindent\strut#1\par%
\vss}\hss}}\fi\@@marginsf\kern-.01cm}
\def\titlepage{%
\frontpagetrue\nopagenumbers\abstractfont%
\hsize=\hstitle\rightline{\vbox{\baselineskip=10pt%
{\abstractfont\pubnum}}}\pageno=0}
\frontpagefalse
\def\pubnum{}
\def\pdate{\number\month/\number\yearltd}
\def\makefootline{\iffrontpage\vskip .27truein
\line{\the\footline}
\vskip -.1truein\leftline{\vbox{\baselineskip=10pt%
{\abstractfont\pdate}}}
\else\vskip.5cm\line{\hss \tenrm $-$ \folio\ $-$ \hss}\fi}
\def\title#1{\vskip .7truecm\titlestyle{\titleft #1}}
\def\titlestyle#1{\par\begingroup \interlinepenalty=9999
\leftskip=0.02\hsize plus 0.23\hsize minus 0.02\hsize
\rightskip=\leftskip \parfillskip=0pt
\hyphenpenalty=9000 \exhyphenpenalty=9000
\tolerance=9999 \pretolerance=9000
\spaceskip=0.333em \xspaceskip=0.5em
\noindent #1\par\endgroup }
\def\autskip{\ifx\answ\bigans\vskip.5truecm\else\vskip.1cm\fi}
\def\author#1{\vskip .7in \centerline{#1}}

\def\address#1{\ifx\answ\bigans\vskip.2truecm
\else\vskip.1cm\fi{\it \centerline{#1}}}
\def\abstract#1{
\vskip .5in\vfil\centerline
{\bf Abstract}\penalty1000
{{\smallskip\ifx\answ\bigans\leftskip 2pc \rightskip 2pc
\else\leftskip 5pc \rightskip 5pc\fi
\noindent\abstractfont \baselineskip=12pt
{#1} \smallskip}}
\penalty-1000}
\def\endpage{\tenpoint\supereject\global\hsize=\hsbody%
\frontpagefalse\footline={\hss\tenrm\folio\hss}}
\def\ack{\goodbreak\vskip2.cm\centerline{{\bf Acknowledgements}}}
\def\CERN{\address{CERN, Geneva, Switzerland}}
\def\bfone{\relax{\rm 1\kern-.35em 1}}
\def\inbar{\vrule height1.5ex width.4pt depth0pt}
\def\IC{\relax\,\hbox{$\inbar\kern-.3em{\mss C}$}}
\def\ID{\relax{\rm I\kern-.18em D}}
\def\IF{\relax{\rm I\kern-.18em F}}
\def\IH{\relax{\rm I\kern-.18em H}}
\def\II{\relax{\rm I\kern-.17em I}}
\def\IN{\relax{\rm I\kern-.18em N}}
\def\IP{\relax{\rm I\kern-.18em P}}
\def\IQ{\relax\,\hbox{$\inbar\kern-.3em{\rm Q}$}}
\def\IR{\relax{\rm I\kern-.18em R}}
\font\cmss=cmss10 \font\cmsss=cmss10 at 7pt
\def\ZZ{\relax\ifmmode\mathchoice
{\hbox{\cmss Z\kern-.4em Z}}{\hbox{\cmss Z\kern-.4em Z}}
{\lower.9pt\hbox{\cmsss Z\kern-.4em Z}}
{\lower1.2pt\hbox{\cmsss Z\kern-.4em Z}}\else{\cmss Z\kern-.4em Z}\fi}
\def\a{\alpha} \def\b{\beta}

 \def\cM{{\cal M}}
 
 \def\cQ{{\cal Q}}

\def\nup#1({Nucl.\ Phys.\ $\us {B#1}$\ (}
\def\plt#1({Phys.\ Lett.\ $\us  {#1}$\ (}
\def\cmp#1({Comm.\ Math.\ Phys.\ $\us  {#1}$\ (}
\def\prp#1({Phys.\ Rep.\ $\us  {#1}$\ (}
\def\prl#1({Phys.\ Rev.\ Lett.\ $\us  {#1}$\ (}
\def\prv#1({Phys.\ Rev.\ $\us  {#1}$\ (}
\def\mpl#1({Mod.\ Phys.\ Let.\ $\us  {A#1}$\ (}
\def\ijmp#1({Int.\ J.\ Mod.\ Phys.\ $\us{A#1}$\ (}
\def\tit#1|{{\it #1},\ }
%

%

\def\tilde{\widetilde}

\def\us#1{\underline{#1}}

\def\Coe#1.#2.{{#1\over #2}}
\def\coeff#1#2{\relax{\textstyle {#1 \over #2}}\displaystyle}
\def\coe#1.#2.{\relax{\textstyle {#1 \over #2}}\displaystyle}

\def\shalf{\relax{\textstyle {1 \over 2}}\displaystyle}

\def\notin{\hbox{{$\in$}\kern-.51em\hbox{/}}}
\def\shdot{\!\cdot\!}
\def\ket#1{\,\big|\,#1\,\big>\,}

\def\exx#1{e^{{\displaystyle #1}}}
\def\del{\partial}

\def\nex#1{$N\!=\!#1$}

\catcode`\@=12
\def\brs{BRST}
\def\qbrs{\cQ_{BRST}}
\def\mm#1#2{\cM_{#1,#2}}
\def\mpq{\mm pq}
\def\ccrit{c^{{\rm(crit)}}}
\def\cpqm{c^{{\rm(matter)}}_{p,q}}
\def\cpql{c^{{\rm(Liouville)}}_{p,q}}
\def\bi#1{b^{[#1]}}
\def\ci#1{c^{[#1]}}
\def\zw#1#2.{{#2\over(z-w)^{#1}}}
\def\tg{T_{gh}}
\def\cW{{\cal W}}
%
\ldf\PolyA{A.\ Polyakov, \plt B103(1981) 207. }
\ldf\LZ{B.\ Lian and G.\ Zuckerman, \plt254B (1991) 417; \plt266B(1991)21.}
\ldf\BMP{P.\ Bouwknegt, J.\ McCarthy and K.\ Pilch, preprints CERN-TH.6162/91
and CERN-TH.6196/91.}
\ldf\Kleb{I.\ Klebanov and A.\ Polyakov, \mpl6(1991) 3273.}
\ldf\PolyB{A.\ Polyakov, \mpl6(1991) 635.}
\ldf\Wit{E.\ Witten, \nup373(1992) 187.}
\ldf\Hull{C.\ Hull, \plt B240 (1990) 110; \nup353(1991) 707;
\plt259B(1991) 68.}
\ldf\Dutch{K.\ Schoutens, A.\ Sevrin and P.\ van Nieuwenhuizen, \cmp124(1989)
87; \plt243B(1990) 245; \plt251B(1990) 355; \nup349(1991) 791. }
\ldf\TM{J.\ Thierry-Mieg, \plt197B(1987) 368.}
\ldf\BLNW{M.\ Bershadsky, W.\ Lerche, D.\ Nemeschansky and N.P.\ Warner,
in preparation.}
\ldf\FLZ{A.B.\ Zamolodchikov, Theor.\ Math.\ Phys.\ 65 (1985) 1205;
V.A.\ Fateev and A.B.\ Zamolodchikov, \nup280(1987) 644;
F.\ Bais, P.\ Bouwknegt, K.\ Schoutens and M.\ Surridge, \nup304(1988) 348;
V.A.\ Fateev and S.L.\ Luk'yanov, \ijmp3(1988) 507.}
\ldf\DK{J.\ Distler and T.\ Kawai, \nup321(1989) 509.}
\ldf\Toda{A.\ Leznov and M.\ Saveliev, \cmp74(1980) 111;
D.\ Olive and N.\ Turok, \nup265(1986) 469; P.\ Mansfield, \nup208(1982) 277;
J.\ Gervais and A.\ Neveu, \nup224(1983) 329;
A.\ Bilal and G.\ Gervais, \nup314(1989) 646; \nup318(1989) 579;
A.\ Bilal, \nup330(1990) 399.}
\ldf\BG{A.\ Bilal and J.\ Gervais, \nup326(1989) 222.}
\ldf\LPSX{H.\ Lu, C.N.\ Pope, S.\ Schrans and K.\ Xu,
preprint CTP-TAMU-5/92, KUL-TF-92/1, KUL-TF-92/1;
H.\ Lu, C.N.\ Pope, S.\ Schrans and X.J.\ Wang, preprint CTP-TAMU-15/92;
C.N.\ Pope, preprint CTP-TAMU-30/92.}
\ldf\MS{P.\ Mansfield and B.\ Spence, \nup362(1991) 294.}
\ldf\KS{Y.\ Kazama and H.\ Suzuki, \plt216B(1989)
112; \nup321(1989) 232.}
\ldf\Moro{A.\ Marshakov, A.\ Mironov, A.\ Morozov and M.\ Olshanetsky, preprint
FIAN/TD-02/92 and ITEP-M-2/92.}
\ldf\Keke{K.\ Li, \nup354(1991) 711.}
\ldf\Ind{S.\ Govindarajan, T.\ Jayamaran, V.\ John and P.\
Majumdar, preprint IMSc-91/40.}
\ldf\Das{S. Das, A. Dhar and S. Kalyana Rama, preprint TIFR/TH/91-20.}
\ldf\Sei{N.\ Seiberg, Progr.\ Theor.\ Phys.\ Suppl.\ 102 (1990) 319.}
\ldf\texans{C.N.\ Pope, L.J.\ Romans and K.S.\ Stelle, \plt268B(1991) 167;
\plt269B(1991) 287.}
\ldf\FMS{D.\ Friedan, E.\ Martinec and S.\ Shenker, \nup271(1986) 93.}
\ldf\wgeo{
C.\ Itzykson, Saclay preprint SPhT/90-174, to appear in:
Proc.\ of Carg\`ese Meeting on Random Surfaces and Quantum Gravity, 1990;
A.\ Bilal, V.\ Fock and I.\ Kogan, \nup359(1991) 635;
A.\ Bilal, \plt249(1990) 56;
A.\ Gerasimov, A.\ Levin and A.\ Marshakov, \nup 360 (1991) 537;
W.\ Sotkov, M.\ Stanishkov and C.J.\ Zhu, \nup356(1991) 245;
W.\ Sotkov and M.\ Stanishkov, \nup356(1991) 439;
J.\ Gervais and Y.\ Matsuo, \plt274(1992) 309 and preprint LPTENS-91-35;
J.\ de Boer and J.\ Goeree, Utrecht preprint THU-92/14.}
\ldf\MSB{P.\ Mansfield and B.\ Spence, preprint NSF-ITP-90-242.}
%


\def\pubnum{
\hbox{CERN-TH.6582/92}
\hbox{HUTP-A034/92}
\hbox{USC-92/015}
\hbox{hepth@xxx/9207067}
}
\def\pdate{
\hbox{CERN-TH.6582/92}
\hbox{July 1992}
}

\titlepage
\title
 {A BRST Operator for non-critical W-Strings}
\author{M.\ Bershadsky}
\address{Lyman Laboratory, Harvard University, Cambridge, MA 02138}
\vskip-.5truecm
\author{W.\ Lerche}
\CERN
\vskip-.5truecm
\author{D.\ Nemeschansky and N.P.\ Warner}
\address{Physics Department, U.S.C., University Park,
Los Angeles, CA 90089}
\vskip-1.truecm
\abstract
{We construct the BRST operator for non-critical $W_3$-strings and discuss the
tachyon-like spectrum. For $N$-punctured spheres with $N \leq 5$
we briefly describe a formal definition of the integral over $W_3$-moduli
space.}

\endpage

\chap{Introduction.}
During the last decade we have learnt much about the structure of
two-dimensional gravity. The gravitational action can be
induced by coupling to the matter system. This coupling is given by
$\delta S=\int d^2z \sqrt g \delta g^{ab}T_{ab}(z)$, where $T_{ab}(z)$ is
the stress-energy tensor of the matter. If this matter
system is conformal, then gravity is governed by the Liouville action \PolyA.
The fact that two-dimensional gravity is invariant under diffeomorphisms
implies the existence of a nilpotent \brs\ charge \FMS
$$
\qbrs\ =\ \oint\!d\!z\,c(z) [T_{zz}^{{\rm(matter)}}+T_{zz}^{{\rm(Liouville)}}+
\shalf T_{zz}^{{\rm(ghost)}}  ]\ ,\eqn\Tbrs
$$
and the physical states are given by the non-trivial \brs\ cohomology. The
$c<1$ conformal minimal models coupled to two-dimensional gravity can have
non-trivial cohomology at any ghost number \doubref\LZ\BMP. For $c=1$, one has
non-trivial cohomology at a finite set of ghost numbers. The states with
$q_{gh}=0$ and $h=0$ make up the ground ring \Wit, while states with $q_{gh}=0$
and $h=1$ generate the symmetry of 2d quantum gravity \nrf\Kleb
\refs{\Wit{--}\PolyB}. This symmetry group, $W_\infty$, is responsible for the
solvability of the system.

 If, on the other hand, the conformal matter system has a larger symmetry then
it can be coupled to some extended background geometry. Thus, in addition to
the well-known generalizations to super-geometry, one can also try to construct
a ``W-geometry'' \wgeo\ in which $W$-gravity is coupled to conformal
$W$-matter \multref\Hull\Dutch\texans. One expects that such a theory is
governed by a
Toda system that extends the Liouville theory of two-dimensional gravity (see,
for example, \Toda). In a sense, this generalization consists in replacing
$SL(2)$ by some other semi-simple Lie group $G$\foot{In this letter we will
mainly consider $G=SL(n)$ and we will take ``$W_n$-gravity'' to refer to this
choice of $G$.}. This amounts to introducing spin $s$ ``gravitons'', where the
spins, $\{s\}$, are given by the degrees of independent Casimirs of $G$ (for
$G=SL(n)$ one has $\{s\}=\{2,3,\dots,n\}$), and leads to what might be called
``$W$-strings'' (see, for example, \multref\BG\Das\LPSX).

In \TM\ a \brs\ charge was constructed for {\it critical} $W_3$-string theory,
that is, for pure $W_3$-matter; nilpotency of $\qbrs$ requires that the matter
central charge must satisfy: $c_M=100$. More generally, for $W_n$ one requires
$$
c_M\ =\ \ccrit \ \equiv\ 4n^3 - 2n -2\ ,\ \qquad  n=2,3,\dots\ . \eqn\crit
$$
The value of $\ccrit$ just balances the contribution from the ghosts $(\bi
i,\ci i), i=1,2,\dots,n-1$ with spins $(i+1,-i)$. The foregoing values of $c_M$
correspond to critical $W_n$-strings, where the Liouville-like degrees of
freedom are expected to decouple. Matter theories with these central charges
can indeed be constructed as described in \LPSX, but these theories are
slightly artificial in that the $W$-generators are constructed from the stress
tensor of the matter by folding in a collection of auxiliary scalars.

On the other hand, the prototype $W_n$-matter theories are
the $W_n$-minimal models, $\mpq$, with central charges \FLZ
$$
\cpqm\ \equiv\ c(\mpq)\ =\ (n-1)\,\Big(1-n(n+1){(p-q)^2\over p q}\Big)\ .
\eqn\cpq
$$
It has been noted \doubref\BG\MSB\ that there is a natural pairing of
$W_n$-Liouville (Toda) theory (denoted by $\mm p{-q}$) with these matter models
in the following sense:
$$
\cpqm\ + \cpql\ =\ \ccrit
$$
(here $\cpql\equiv c(\mm p{-q})$). These theories appear to be dual in the
sense described in \LZ, and this suggests that the tensor product models
$$
\cW^{(n)}_{p,q}\ \equiv\
\mpq^{(W_n {\rm matter)}}\otimes\mm p{-q}^{(W_n {\rm Liouville)}} \otimes \{\bi
i,\ci i\}\eqn\tensor
$$
describe non-critical $W_n$-strings \MS.  These models are
conceivably solvable and are thus, for us, the most interesting ones.
A further observation \doubref\BG\MSB\  is that there is a similar natural
pairing in the construction of tachyonic states: if $V_{r_i,s_i}$ denotes
a vertex operator in either the Liouville or matter sector in the usual
parametrization, then
$$
h(V_{r_i,-s_i}^{{\rm(Liouville)}}) + h(V_{r_i,s_i}^{{\rm(matter)}})\ =\
h_{{\rm crit}}\ \equiv\ \coeff16 n(n^2-1)\ \ \ \ \ \forall r_i, s_i\ .
\eqn\hadd
$$
This is just the maximal dimension that can
be compensated by all of the $c$-ghosts, and suggests
that at least the following, tachyonic operators should be physical:
$$
\eqalign{
&T_{r_i,s_i}\ =\ X(c)\,
V_{r_i,-s_i}^{{\rm(Liouville)}}V_{r_i,s_i}^{{\rm(matter)}}\ ,\qquad\qquad
h(T_{r_i,s_i})\equiv0\ ,\cr
&X(c)\ \equiv\ \prod_{i=1}^{n-1}[\ci i\del\ci i\del^2\ci i\dots\del^{i-1}\ci
i]\ .}\eqn\tach
$$

While these observations, and some other pieces of circumstantial evidence,
suggest a natural formulation of non-critical $W$-strings, the construction of
a proper \brs\ charge has been lacking. Without such a charge, the discussion
of the physical states can only be rather speculative. The problem in
constructing the \brs\ charge lies in the non-linearity of the $W$-algebra, and
so the $W$-matter models apparently cannot be coupled to $W$-gravity using the
\brs\ operator of \TM.

The main purpose of this letter is to construct explicitly the \brs\ charge for
$W_3$-matter coupled to $W_3$-gravity, thus demonstrating that this tensor
product actually does yield a sensible theory (and strongly suggesting that
this is true for all $n$). Furthermore, we show that the tachyonic operators
\tach\ are indeed \brs\ invariant, as expected. We also discuss some issues
related to $W$-moduli. In addition, we make some brief remarks on the extra
states of the associated $c_M=2$ model (corresponding to $p=q$ above).

\chap{The BRST operator}

The construction of $\qbrs$ for $W_3$-Liouville coupled to $W_3$-matter
proceeds in the simplest possible manner and closely parallels the computation
of \TM. Following  \FLZ\ we take
$$
\eqalign{
W(z)W(w)\ &=\ \zw6c/3.+\zw42T(w).+\zw3\del T(w).\cr
&+ \zw21.[2b^2\Lambda(w)+\coeff3{10}\del^2T(w)]
+\zw{}1.[b^2\del\Lambda+\coeff1{15}\del^3T(w)]\cr
&+ {\rm regular\ terms}\ ,}
\eqn\walg
$$
where $c$ is the central charge and $b^2\equiv{16\over5c+22}$. The composite
operator on the right-hand side is defined by $\Lambda(z)= (TT)(z) -
\coeff3{10}\del^2T(z)$. (Normal ordering of two operators $A(z)$ and $B(z)$ is
defined by $(AB)(z)={1\over2\pi i}\oint_z d\zeta\coeff1{\zeta-z}A(\zeta)B(z)$.)
Now consider a theory that consists of a tensor product of $W_3$-matter coupled
to $W_3$-Liouville\foot{Note that our construction of a \brs\ operator will
only assume the operator product structure of the $W_3$-algebra, and is
therefore not restricted merely to the models $\cW^{(3)}_{p,q}$ in \tensor.},
and let
$W_L(z)$, $W_M(z)$, $T_L(z)$, $T_M(z)$, $c_L$ and $c_M$ denote the currents and
central charges. The stress tensor of the $j$-th ($j=1,2$) ghost system
is given by:
$
\tg^{[j]}=-(j+1)\bi j(\del\ci j)-j(\del\bi j)\ci j,
$
which has central charge $c_j=-2(6j^2+6j+1)$.  The total ghost
contribution to the central charge is thus $c_{gh}=-100\equiv-\ccrit$.

We find that the \brs\ current $J(z)$ has the form
(up to total derivatives):
$$
\eqalign{
J(z)\ &=\ \ci2(z)\big[\tilde W_L(z) \pm i \tilde W_M(z)\big]
+ \ci1(z)\big[T_L(z)+T_M(z)+\shalf\tg^{[1]}(z)+\tg^{[2]}(z)\big]\cr
&+ \big[T_L(z)-T_M(z)\big]\bi1(z)\big(\ci2\del_z\ci2(z)\big)
+ \mu\big(\del_z\bi1(z)\big)\ci2(z)\big(\del^2_z\ci2(z)\big)\cr
&+ \nu\bi1(z)\big(\ci2\del^3_z\ci2(z)\big)\ ,
}\eqn\jbrs
$$
where $\tilde W={1 \over b}W$ and $\mu={3 \over 5}\nu= {1 \over 10{b_L}^2}
(1-17{b_L}^2)$. The choice of the first two terms is rather natural, and the
form of the entire current, $J(z)$, is similar to that of \TM. To calculate
$(\qbrs)^2$, where $\qbrs = \oint J(z) dz$, one computes the simple pole terms
in $J(z)J(w)$, and discards all total derivatives. Rather than recount the
calculation in detail we simply note the following features of the calculation:
\item{(i)}The contribution of the operator product of the second term,
$\ci1[T_L+T_M+\dots]$, with itself is the standard \brs\ computation and gives
rise to the condition $c_L+c_M=100$.
\item{(ii)}Essentially because the first term is a primary field of weight one,
its operator product with the second term yields simple pole terms that are
purely total derivatives.
\item{(iii)}The third term has been chosen so that its operator product with
the second term gives rise to a factor of the form: $\zw{}1.[(T_LT_L)-(T_MT_M)]
\ci2\del\ci2$, which cancels against a similar term arising from
$\ci2[\tilde W_L \pm i\tilde W_M](z)\ast\ci2[\tilde W_L \pm i \tilde W_M](w)$.
\item{(iv)} The coefficients $\mu$ and $\nu$ are (over)determined by the
requirement that all simple pole terms proportional to $(\del^mT)(\del^n\ci2)
(\del^p\ci2)$ are total derivatives.
\item{(v)}The many other terms that arise in computing $(\qbrs)^2$ then vanish
as a consequence of the choice of coefficients and the requirement
that $c_L+c_M=100$.

\noindent
As a result of the computation, we find that $J(z)$ yields a nilpotent \brs\
charge. The only ambiguity is the choice of sign in front of $i\tilde W_M$. In
fact, the \brs\ current is almost invariant under the interchange of the matter
and Liouville systems, modulo conjugation. That is, rescaling the ghosts
$c^{[2]} \rightarrow \mp i c^{[2]}$ and $b^{[2]} \rightarrow \pm i b^{[2]}$ one
effectively flips the matter and Liouville systems, provided that one has:
$$
{1 \over 10b_L^2}(1-17b_L^2)=-{1 \over 10b_M^2}(1-17b_M^2)\ .
$$
However, this identity is equivalent to $c_L+c_M=100$.

\chap{Physical states}
We now look for physical, tachyonic states that are the counterparts
of the Distler-Kawai states \DK\ of ordinary matter and gravity.
Specifically, we seek physical states of the form
$$
\ket v\ =\ \ket{v_L}\otimes\ket{v_M}\otimes\ket{\tilde0}_{gh}\ ,\eqn\pstat
$$
where $\ket{v_L}$ and $\ket{v_M}$ are pure momentum states in the Liouville
and matter sectors, and $\ket{\tilde0}_{gh}$ is the ghost vacuum with
$$
\eqalign{
\ci j_n\ket{\tilde0}_{gh}\ &=\ 0\ , \qquad n\geq1\cr
\bi j_n\ket{\tilde0}_{gh}\ &=\ 0\ , \qquad n\geq0\cr
(L_0)_{gh}\ket{\tilde0}_{gh}\ &=\ -h_{{\rm crit}}\ket{\tilde0}_{gh}\ =\
-4\ket{\tilde0}_{gh}\ .}\eqn\ghvac
$$
If $\ket0$ denotes the $SL(2,\IR)$ invariant ghost vacuum with
$\ci j_n\ket{0}= 0,n\geq(j+1),\ \bi j_n\ket{0}= 0,n\geq-j$,
then $\ket{\tilde0}_{gh}=\ci2(0)\del\ci2(0)\ci1(0)\ket0\equiv X(c)\ket0$.
On states of the form \pstat, one finds that the physical state condition
$\qbrs\ket v=0$ reduces to
$$
h_L + h_M\ =\ 4\ \ ,\quad\qquad \tilde w_L\pm i\tilde
w_M\ =\ 0\ .\eqn\physcond
$$
Here, $h$ and $\tilde w$ are the eigenvalues of
$L_0$ and $\tilde W_0\equiv\coeff1bW_0$, respectively.
The $\pm$ sign in \physcond\ corresponds to the choice of sign in
\jbrs, and henceforth we will choose the {\it positive} root.
To compute these eigenvalues we introduce free field realizations of the
currents:
$$
\eqalign{
T_M(z)\ &=\ -\shalf(\del\phi_M)^2 - i\a_0\, \rho\shdot\del^2\phi_M \cr
T_L(z)\ &=\ -\shalf(\del\phi_L)^2 - \b_0\, \rho\shdot\del^2\phi_L\ ,
}\eqn\TML
$$
where $\phi_M(z)$ and $\phi_L(w)$ are both two-component vectors of bosons with
$\phi_M^a(z)\phi_M^b(w)$ $\sim \phi_L^a(z)\phi_L^b(w) \sim-\delta^{ab}
ln(z-w)$. The vector, $\rho$, is the Weyl vector of $SU(3)$, and satisfies
$\rho^2=2$. The background charge parameters, $\a_0$ and $\b_0$, are real, and
the condition $c_L+c_M=100$ is equivalent to ${\b_0}^2-{\a_0}^2=4$. One can
parameterize such a $\b_0$ and $\a_0$ by introducing a real, positive
parameter, $t$, and setting $\a_0=\sqrt t-1/\sqrt t$ and $\b_0=\sqrt t+1/\sqrt
t$. (For the models $\cW^{(3)}_{p,q}$ in \tensor\ one has $t = q/p$.) In this
representation, the $W$-generator takes the form \FLZ
$$
\eqalign{
\tilde W(z)\ &=\
-{i\over12}\big[(\del\psi_2)^3-3(\del\psi_1)^2(\del\psi_2)+3\gamma
(\del^2\psi_1)(\del\psi_2)\cr
&\ \ \ + 9\gamma(\del\psi_1)(\del^2\psi_2)-6\gamma^2(\del^3\psi_2)\big]\ ,}
\eqn\wgen
$$
where $\tilde W(z) \equiv \coeff1b W(z)$. For the matter sector one
takes $\gamma=i\a_0=i(\sqrt t-1/\sqrt t)$ and
$\psi_1=-(\a_1+\a_2)\shdot\phi_M$, $\psi_2=\coeff1{\sqrt3}
(\a_1-\a_2)\shdot\phi_M$, where $\a_1$ and $\a_2$ are the simple roots of
$SU(3)$. For the Liouville sector one takes $\gamma=\b_0=\sqrt t+1/\sqrt t$ and
$\psi_1=-(\a_1+\a_2)\shdot\phi_L$, $\psi_2=\coeff1{\sqrt3}
(\a_1-\a_2)\shdot\phi_L$. Introduce vertex operators
$$
\eqalign{
V_M(a_1,a_2)\ &=\ \exp[i(a_1\lambda_1+a_2\lambda_2)\shdot\phi_M]\cr
V_L(b_1,b_2)\ &=\ \exp[(b_1\lambda_1+b_2\lambda_2)\shdot\phi_L]\ ,}
\eqn\vertex
$$
where $\lambda_{1,2}$ are the fundamental weights of $SU(3)$ with
$\a_i\shdot\lambda_j=\delta_{ij}$. From \wgen\ and \TML, a simple computation
shows that for these vertex operators one has:
$$
\eqalign{
h_M(a_1,a_2)\ &=\ \coeff1{12}[3(a_1+a_2+2\a_0)^2+(a_1-a_2)^2-12{\a_0}^2]\cr
\tilde w_M(a_1,a_2)\ &=\
-\coeff1{9\sqrt3}(a_1-a_2)(2a_1+a_2+3\a_0)(a_1+2a_2+3\a_0)\cr
h_L(b_1,b_2)\ &=\ -\coeff1{12}[3(b_1+b_2+2\b_0)^2+(b_1-b_2)^2-12{\b_0}^2]\cr
\tilde w_L(b_1,b_2)\ &=\ -\coeff
i{9\sqrt3}(b_1-b_2)(2b_1+b_2+3\b_0)(b_1+2b_2+3\b_0)\ .}\eqn\eigenv
$$
{}From this it is easy to see that if $b_1=a_1+\a_0-\b_0$
and $b_2=a_2+\a_0-\b_0$ then the physical state condition \physcond\
is satisfied.  There are, however, further solutions to \physcond\ that
can be generated by the action of the Weyl group on the foregoing
obvious solution.   A simple way to see this is to introduce the
orthonormal basis: $e_i$, $i = 1,2,3$, in which the simple roots
can be written $\alpha_1 = e_1 -e_2$, $\alpha_2 = e_2 -e_3$.  Now
consider the vector
$$
u \equiv b_1 \lambda_1 ~+~ b_2 \lambda_2 ~+~ \beta_0 \rho ~=~
\coeff13(2b_1 + b_2 + 3 \beta_0) e_1 -  \coeff13(b_1 - b_2) e_2 -
\coeff13(b_1 + 2b_2 + 3\beta_0)e_3 \ ,
$$
and observe that $\tilde w_L(b_1,b_2)$ is a constant multiple of
the product of the components of $u$.  Moreover, one has
$h_L(b_1,b_2) = - \coeff12 u^2 + \beta_0^2$.  Since the Weyl
group acts by permuting the $e_i$, it follows that the vertex
operator $\exp[(\sigma(b_1\lambda_1+b_2\lambda_2 + \beta_0 \rho)
- \beta_0 \rho)\shdot\phi_L]$ has the same values of $h_L$ and
$\tilde w_L$ for any element $\sigma$ of the Weyl group of $SU(3)$.
It follows that the operators
$$
T_{\Lambda,\sigma} (z)\ =\ \ci2(z)\del\ci2(z)\ci1(z)\, \exp\!\big[
i \Lambda\shdot\phi_M(z) + (\sigma(\Lambda + \alpha_0 \rho) -
\beta_0 \rho) \shdot \phi_L(z) \big], \eqn\physvert
$$
for any $\Lambda$, and for any choice of $\sigma$ in the Weyl group of
$SU(3)$, create physical states on the $SL(2,\IR)$ invariant
vacuum of the theory. Note that because of the Weyl rotations in
\physvert, each matter field has six possible Liouville dressings. We expect
that there exists some analogue of Seiberg's condition \Sei\ that selects
one choice of dressing. In particular it seems reasonable that if $\Lambda$
is in the fundamental Weyl chamber, then one should take $\sigma=1$. This
corresponds to the operators \tach\ mentioned in the introduction.

\chap{Additional comments}

In analogy with two-dimensional gravity \LZ, we expect that there will be other
physical states in addition to the tachyons \physvert. In particular, we
anticipate that there will be physical states with different ghost numbers. The
$W$-minimal models coupled to $W$-gravity clearly should have physical states
for any ghost numbers, while the physical states of the $c=2$ theory will
appear at a finite set of ghost numbers\foot{The $c_M=2$ model corresponds to a
very intriguing $4$-dimensional string theory. The target space has $(2,2)$
signature and Lorentz symmetry is broken by the dilaton background.}. The
operators at different ghost numbers might be constructed by using the \brs\
current \jbrs\ and the corresponding descent equations similar to those
employed in \doubref\Wit\Ind. The whole structure will be more complicated than
for ordinary gravity precisely because of the larger number of extra states and
more involved structure of the descent equations. The ghost number zero,
dimension zero operators should certainly form a ground ring. We will discuss
these states and ground rings in a later paper \BLNW, and here we will simply
make some brief remarks about the theory with $c_M=2$ at the $SU(3)$ symmetric
point.

First observe that $c_M=2$ corresponds to $\a_0=0$.
We can therefore introduce the $SU(3)$ currents
$$
J^{\pm\a_j}(z)\ =\ \exx{\pm i\a_j\shdot\phi_M(z)}\ ,\eqn\sucur
$$
where $\a_j, j=1,2,3$ are the positive roots of $SU(3)$. One can verify that
the zero modes of these currents commute with $T_M(z)$ and $\tilde W_M(z)$.
Consequently, if $\Lambda$ is a weight of $SU(3)$, we may use the generators
\sucur\ on a tachyonic state $T_{\Lambda,\sigma}(z)$ so as to obtain an entire
$SU(3)$ multiplet of physical states of the $W$-string. Note that the
``corners'' of a weight diagram will correspond to tachyon states \physvert,
while the rest of the multiplet corresponds to ``extra states'' that are
descendents of vertex operator states. Such $SU(3)$ discrete states have
already been discussed in \Moro. We therefore find that this part of the
structure of $W_3$-gravity coupled to $W_3$-matter is a relatively
straightforward generalization of $c=1$ matter coupled to ordinary gravity.

Having remarked upon the similarities between $W$-strings and ordinary strings,
we now would like to note some fundamental differences. The basic problem now
is how to construct correlation functions, or more generally, how to understand
$W$-geometry and the integration over $W$-moduli. For example, the $c^{[2]}$
ghost number anomaly on the sphere is equal to $5$, but three insertions of
$X(c)$ have $c^{[2]}$-ghost charge equal to $6$. This implies that even the
three-punctured sphere has a non trivial $W$-modulus, that is, in contrast to
usual gravity, the sphere is not ``rigid'' in $W$-gravity. The $N$-punctured
sphere has $(N-3)+(2N-5)$ moduli, and therefore a $N$-point correlation
function should involve additional $2N-5$ integrations over the $W$-moduli. One
cannot thus employ only the standard tachyons \physvert\ to make non-vanishing
correlation functions in $W_n$-models for $n>2$; one also needs physical
operators at different ghost numbers. In analogy to ordinary strings, we expect
that each physical state can be represented in different ``$W$-pictures'' at
various ghost numbers. Such pictures can probably have all possible
permutations of ghost content at each fixed ghost number.

We suggest the following partial resolution of this problem.
Let us first define the total stress-energy tensor and the total spin-$3$
current\foot{${\cal T}$ and ${\cal W}$ do not appear to generate a
$W$-algebra.}
$${\cal T}=[Q_{BRST}, b^{[1]}]~,~~~{\cal W}=[Q_{BRST}, b^{[2]}].
$$
Each tachyonic vertex operator \physvert\ can be mapped to representatives, or
``avatars'', in other $W$-pictures. Each tachyon will have avatars with at
least ghost numbers $0,1,2$ and $3$. Here we present only those avatars with
ghost numbers $1,2$ and $3$ that are relevant for the discussion below:
$$
\eqalign{&\Phi_\Lambda^{(3)}=c^{[1]} c^{[2]} \partial c^{[2]} V_\Lambda
\ \equiv T_\Lambda~,\cr
\Phi^{(2)}_{\Lambda,1} = &c^{[2]} \partial c^{[2]} V_\Lambda ~~,~~~~
\Phi^{(2)}_{\Lambda, 2}=c^{[1]} \partial c^{[2]} V_\Lambda + \dots\cr
&\Phi^{(1)}_\Lambda = \partial c^{[2]} V_\Lambda + \dots\ .\cr }
\eqn\pict
$$
The operators at different ghost numbers are related to each other by
descent equations similar to those employed in ordinary gravity:
$$
\eqalign{&[Q_{BRST}, \Phi^{(3)}]=0~,\cr
&[Q_{BRST}, \Phi^{(2)} _1]={\cal L}_{-1} \Phi^{(3)}~,\cr
&[Q_{BRST}, \Phi^{(2)} _2]=-{\cal W}_{-2} \Phi^{(3)}~,\cr
&[Q_{BRST}, \Phi^{(1)}]={\cal W}_{-2} \Phi^{(2)} _1 +
                {\cal L}_{-1} \Phi^{(2)} _2~.\cr}
\eqn\desc
$$
These descent equations should involve only ${\cal W}_{-2}$ and ${\cal
L}_{-1}$. The complete structure of $W$-avatars is more complicated. It is
crucial that ${\cal W}_{-2}$ and ${\cal L}_{-1}$ commute with each other and
therefore they may be realized as derivations: ${\cal L}_{-1}=\partial_z$ and
${\cal W}_{-2}=\partial_{\xi}$. It is natural to define a ``$W$-field'' as
follows:
$$
{\cal P}(z, \xi)=e^{\xi {\cal W}_{-2}}\, \Phi(z)\ .
$$
The new coordinate $\xi$ corresponds to the $W$-modulus. One can also define
${\cal P}^{(3)}, {\cal P}^{(2)}_1 , {\cal P}^{(2)}_2$ and
${\cal P}^{(1)}$, which are related to each other by descent equations \desc.
Now it is almost obvious that
$$\int\!d\xi\, {\cal P}^{(2)} _2 (z, \xi)~~,~~~~
\int\!d \xi dz\, {\cal P}^{(1)}(z, \xi)
$$
are (at least formally) \brs\ invariant operators\foot{Note also that
$z$-integrals over the screening operators are \brs-invariant. These operators
for the Liouville sector (which appear in the Toda action) are thus the
simplest examples for discrete states with (for $W$-gravity) non-standard
ghost number.}. Using these operators
one can construct 2-, 3-, 4- and 5-point correlation functions. The
failure of this approach to yield arbitrary $N$-point correlation
functions is due to the complicated structure of the $W$-moduli for
$N>5$. To construct the general correlation functions one has to utilize
the properties of the complete collection of $W$-avatars. We intend to discuss
the properties of $W$-avatars in one of our next publications.

Finally, we would like to point out that it is the appearance of a holomorphic
structure that is responsible for various remarkable properties of this class
of theories. For example, the fact that the dimension of the tachyon operator
\physvert\ vanishes for all $\Lambda$, rests crucially on the
holomorphic\foot{With ``holomorphic'' we mean holomorphic up to Weyl
transformations $\sigma$ in \physvert. We can always choose a representative in
each Weyl orbit such that only the holomorphic combination $\Lambda \shdot
(\phi_L+i\phi_M)$ appears.} combination $\Lambda \shdot
(\phi_L+i\phi_M)$. Similarly, physical states satisfy the holomorphic condition
$\tilde w_L+ i\tilde w_M = 0$. These features are reminiscent and
actually related to similar properties of topological, twisted \nex2
superconformal theories. Indeed, one can show that for the topological models,
$\cW^{(n)}_{p=1,q=n+k}$, the spectrum contains operators that have an algebraic
structure analogous to the chiral primary fields of \nex2 coset models of \KS\
based on $SU(n)_k/U(n-1)$. More specifically, for these models the tachyons
\tach\ can be viewed as $T_\Lambda = e^{\Lambda \cdot\Phi}P$, where
$\Phi=(\phi+i\varphi)$ plays the role of the bottom component of a chiral
superfield and its exponentials correspond to a free superfield realization of
the primary chiral fields of the coset models. The remaining non-holomorphic
piece, $P$, is the $W$-gravity puncture operator. Similarly, the ground rings
of ghost-neutral operators seem to contain the chiral rings of the coset
models, together with their $W$-gravitational extensions. Thus, the models
$\cW^{(n)}_{1,n+k}$ seem also to describe topological $W_n$-gravity coupled to
topological minimal $W_n$-matter models at level $k$, in correspondence to the
results (for $n=2$) of \Keke. We will present a detailed analysis of the
various ring structures in \BLNW.

\ack

We wish to thank  P.\ Bouwknegt, K.\ Li, K.\ Pilch,    A.\ Sevrin  and H.\
Verlinde for discussions. M.B.\ is partially supported by NSF grant PHY
87/14654 and by Packard Fellowship 89/1624, D.N.\ and N.P.W.\ are
supported in part by funds provided by the DOE
under grant No. DE-FG03-84ER40168.  N.P.W. is also partially supported
by a fellowship from the Alfred P. Sloan Foundation.

\refout

\end